\documentclass[12pt,letterpaper]{article}
\usepackage[top=0.85in,left=0.75in,footskip=0.75in,marginparwidth=2in]{geometry}

\usepackage[utf8]{inputenc}

\usepackage{nameref,hyperref}

\usepackage[right]{lineno}

\usepackage{microtype}
\DisableLigatures[f]{encoding = *, family = * }

\raggedright
\usepackage{changepage}

\usepackage[aboveskip=1pt,labelfont=bf,labelsep=period,singlelinecheck=off]{caption}

\makeatletter
\renewcommand{\@biblabel}[1]{\quad#1.}
\makeatother

\usepackage{lastpage,fancyhdr,graphicx}
\usepackage{epstopdf}
\usepackage{color}

\definecolor{Gray}{gray}{.25}

\usepackage{graphicx}

\usepackage{sidecap}

\usepackage{wrapfig}
\usepackage[pscoord]{eso-pic}
\usepackage[fulladjust]{marginnote}
\reversemarginpar

\usepackage{mathtools}

\usepackage{algorithm}
\usepackage[noend]{algpseudocode}

  \usepackage[
    backend=biber,
    style=authoryear,
    citestyle=numeric-comp,sorting=none,
    natbib=true,
    firstinits=true,uniquename=false,uniquelist=false,maxcitenames=2,
    url=false,labelnumber 
  ]{biblatex}

\DeclareNameAlias{sortname}{last-first}

\AtEveryBibitem{\clearfield{month}}
\AtEveryBibitem{\clearfield{day}}
\AtEveryBibitem{\clearfield{url}}
\AtEveryBibitem{\clearfield{doi}}
\AtEveryBibitem{\clearfield{issn}}
\AtEveryBibitem{\clearfield{urldate}}  
\AtEveryBibitem{\clearfield{language}} 
\AtEveryBibitem{\clearlist{language}}
\AtEveryBibitem{\clearlist{extra}}
\AtEveryBibitem{\clearfield{extra}} 
\AtEveryCitekey{\clearlist{extra}}
\AtEveryBibitem{\clearlist{note}}
\AtEveryBibitem{\clearfield{note}} 
\AtEveryCitekey{\clearlist{note}}
\AtEveryBibitem{\clearlist{issue}}
\AtEveryBibitem{\clearfield{issue}} 
\AtEveryCitekey{\clearlist{issue}}
\AtEveryBibitem{\clearlist{number}}
\AtEveryBibitem{\clearfield{number}} 
\AtEveryCitekey{\clearlist{number}} 

\DeclareFieldFormat{labelnumberwidth}{#1.}

\defbibenvironment{bibliography}
  {\list 
    {\printfield[labelnumberwidth]{labelnumber}} 
  {\setlength{\labelwidth}{\labelnumberwidth}%
   \setlength{\leftmargin}{0pt}%
   \setlength{\labelsep}{\biblabelsep}%
   \setlength{\itemsep}{\bibitemsep}%
   \setlength{\itemindent}{\labelwidth}%
   \addtolength{\itemindent}{\labelsep}%
   \setlength{\parsep}{\bibparsep}}%
   } 
{\endlist} 
{\item}

\renewbibmacro*{in:}{}
\DeclareFieldFormat{pages}{#1}

\usepackage[space]{grffile}

\def\bw{\mathbf{w}}
\def\bx{\mathbf{x}}

\usepackage{amsmath,amssymb}

\newcommand{\Exp}[1]{\left\langle\,{#1}\,\right\rangle}

\newcommand{\appropto}{\mathrel{\vcenter{
  \offinterlineskip\halign{\hfil$##$\cr
    \propto\cr\noalign{\kern2pt}\sim\cr\noalign{\kern-2pt}}}}}

\begin{document}
\vspace*{0.05in}

\begin{flushleft}
{\Large
\textbf\newline{Correlation-invariant synaptic plasticity}
}
\newline
\\
Carlos Stein Brito\textsuperscript{1,2}, 
Wulfram Gerstner\textsuperscript{1}
 \\
 \bigskip
 \textsuperscript{1}École Polytechnique Fédérale de Lausanne, EPFL, Switzerland \\
 \textsuperscript{2}Champalimaud Research, Champalimaud Foundation, Portugal
\end{flushleft}

\justify

\section*{Abstract}
Cortical populations of neurons develop sparse representations adapted
to the statistics of the environment.
While existing synaptic plasticity models reproduce some of the observed
receptive-field properties, a major obstacle
is the sensitivity of Hebbian learning to omnipresent spurious
correlations in cortical networks
which can overshadow relevant latent input features. Here we develop a
theory for synaptic plasticity
that is invariant to second-order correlations in the input. Going
beyond classical Hebbian
learning, we show how Hebbian long-term depression (LTD) cancels the
sensitivity to second-order
correlations, so that receptive fields become aligned with features
hidden in higher-order statistics.  Our simulations demonstrate how
correlation-invariance enables biologically realistic models to develop
sparse population codes, despite diverse levels of variability and
heterogeneity.
The theory advances our understanding of local unsupervised learning in
cortical circuits and
assigns a specific functional role to synaptic LTD mechanisms in
pyramidal neurons.


\section*{Introduction}

Sensory networks contain rich representations of the external world, with individual neurons responding selectively to particular stimuli \citep{hubel_receptive_1959, desimone_face-selective_1991}. These representations develop in early life and continue onward to adapt to the statistics of the environment \citep{goldstone_perceptual_1998}. While synaptic plasticity is thought to be central to cortical learning, it is still unknown how these biological processes learn complex representations. 

Models of excitatory synaptic plasticity can reproduce some of these findings, and support the development of sparse codes from natural stimuli, but rely on unrealistic assumptions of decorrelated inputs and identical firing rates or an algorithmic whitening step \citep{olshausen_emergence_1996,clopath_connectivity_2010, zylberberg_sparse_2011}. In these simplified settings, many plasticity models display similar behaviour and develop the expected range of receptive fields \citep{rehn_network_2007,brito_nonlinear_2016}. However, the relation of plasticity rules derived from sparse-coding models to experimental data remains often at a high level and cannot explain differences between homosynaptic LTD on one side and neuron-wide (heterosynaptic) depression mechanisms or homeostasis on the other side \citep{turrigiano_homeostatic_2004, zenke_hebbian_2017, wu_homeostatic_2020}.

Here we develop a theory of sparse feature learning which takes into account the diverse statistics of presynaptic neurons, such as noise, correlations and heterogeneous firing rates \citep{barth_experimental_2012}. 
We demonstrate that invariance to input correlations requires plasticity models to have nonlinear Hebbian LTP and standard Hebbian LTD, linked with a homeostatic factor of meta-plasticity, which includes variations of the BCM \citep{cooper_bcm_2012} and the triplet STDP models \citep{pfister_triplets_2006}, classic models of excitatory plasticity, as special cases. We show that this family of plasticity models optimizes an objective function, similar to that of 
sparse coding models \citep{foldiak_forming_1990,olshausen_emergence_1996}, but with the additional constraint of invariance to second-order correlations. Thus our objective function aims to detect sparse features while ignoring potentially large second-order correlations in the synaptic input.

Our simulations demonstrate how correlation-invariance enables biologically realistic models to learn efficient decoders and sparse population codes. Applied to sensory integration tasks, optimizing for sparsity translates to optimal integration of noisy inputs, weighing them according to their scale and reliability, leading to near-optimal linear decoders. In connected populations of neurons, the same plasticity rule leads to precisely tuned neurons even in cases where inputs have strong spatial correlations. In a spiking model of sensory development, we show that correlation-invariant STDP is sufficient to learn localized receptive fields from natural images, while models without this property require alternative mechanisms to promote sparsity, such as lateral inhibition.

Correlation-invariant learning assigns a functional role to LTP, LTD, and homeostasis. In particular, linear Hebbian LTD is critical for stable learning, whereas alternative stability mechanisms, such as heterosynaptic plasticity \citep{turrigiano_homeostatic_2004, zenke_diverse_2015}, do not confer correlation-invariance. Our theory of correlation-invariant learning provides a normative explanation for the existence of several distinct plasticity mechanisms in the brain. These results extend our understanding of how unsupervised learning with local Hebbian plasticity might be implemented in cortical circuits.

\section*{Results}

\subsection*{Synaptic plasticity as sparse feature learning}

We hypothesize that synaptic plasticity in single neurons implements an algorithm to learn features hidden in the input arriving in parallel at multiple synapses. 
In this view, the formation of receptive fields of sensory neurons during development is a manifestation of successful feature learning.

We start by considering a simplified rate neuron $y$, with activation $y = (\bw^T \bx)_+$, receiving $N$ inputs $\bx = (x_1, \hdots, x_N)$ through synaptic connections $\bw = (w_1, \hdots, w_N)$, where $(.)_+$ denotes the rectified linear activation function, with activity $y=\bw^T \bx$ for $\bw^T \bx>0$ and $y=0$ otherwise. We assume that input features are characterised by sparse, non-Gaussian, statistics, as in sparse coding and independent component analysis (ICA) frameworks \citep{olshausen_emergence_1996,bell_independent_1997}.
 
It is possible \citep{oja_learning_1991, savin_independent_2010, zylberberg_sparse_2011} to learn such features with local plasticity models provided the inputs have been decorrelated and normalized, i.e. whitened, by having been preprocessed to have an identity covariance matrix and unit firing rates. For such preprocessed inputs, it has been shown that a large class of sparsity maximization methods can retrieve the latent features \citep{brito_nonlinear_2016}. Classically the sparseness of the output activity is measured through an objective function $\langle F(y) \rangle$, where $\langle . \rangle$ denotes the expectation over the data samples $\{\bx\}$  \citep{oja_learning_1991,hyvarinen_independent_1998}. An online plasticity rule (derived e.g. via stochastic gradient descent) converges to a solution that maximizes this objective:
\begin{equation}
\left.
\begin{aligned}
(\text{1-a}) \ \ \Delta \bw  & = \eta \ \bx \ f(y) \\
(\text{1-b})  \ \ \ \ \bw & \gets \frac{\bw + \Delta \bw}{\|\bw + \Delta \bw\|} 
\end{aligned}
\ \
\right\}
\xRightarrow[\eta \to 0]{converges} \ \bw = \text{argmax}_{\bw, |\bw| = 1} \ \langle F(y) \rangle 
  \label{nhl}
\end{equation}
where $\eta$ is a learning rate and $f(.)$ is the derivative of $F(.)$.  
In particular, if $F(y) = \frac{1}{3} y^3$ then $f(y) = y^2$, which relates to known experimental and theoretical results for activity-dependent models, as discussed below. 
The learning rule of Eq.\ref{nhl}-a can be interpreted as a model of activity-dependent synaptic plasticity with a nonlinear Hebbian form of LTP. Eq.\ref{nhl}-b assures normalization of the weight vector and can be related to weight decay \cite{miller_role_1994}. 

Normalization is a strict form of stabilization of the weight vector. A weaker form of stabilization can be achieved through dynamical mechanisms, such as heterosynaptic depression \citep{zenke_diverse_2015}. However, if different input neurons have diverse firing rates or correlations between them, the simple sparsity objectives and related learning rules mentioned above do not learn the desired features. Instead of retrieving sparse features, they learn the input directions of the largest variance, as do PCA methods.

\subsection*{Theory of correlation-invariant learning}

We aim for a synaptic plasticity rule that is capable of extracting low-amplitude features even if synaptic inputs exhibit spurious correlations of large amplitude. Here spurious refers to modulations with a Gaussian amplitude distribution whereas features are defined by a sparse non-Gaussian distribution. 
As shown in Methods, an online update rule with LTP and LTD solves the \emph{correlation-invariant} optimization problem $\Exp{F(y)} = \Exp{\left(\frac{y}{\sigma_y}\right)^3}$
in a rectified linear neuron $y = (\bw^T \bx)_{+}$:
\begin{equation}
\left.
\begin{aligned}
(\text{2-a}) \ \ \ \Delta \bw & = \eta \ ( \bx \ y^2 - h_y \ \bx \ y)  \\
(\text{2-b}) \ \ \ \Delta h_y & = \eta_h \ (y^2 - h_y)
\end{aligned}
\ \
\right\}
\xRightarrow[\eta \to 0]{converges} \ \bw = \text{argmax}_{\bw} \ \Exp{ \left(\frac{y}{\sigma_y} \right)^3 }
  \label{invrule}
\end{equation}
Importantly, weight vectors are not constrained to norm one, but the output
activity is normalized by its standard deviation, $\sigma_y = \sqrt{\Exp{y^2}}$. We define \emph{correlation-invariant} objectives as being invariant to the input correlations, and consequently invariant to linear transformations of the input such as rescaling or whitening, as demonstrated in Methods.

We note that, while invariant to correlations, this sparsity objective is still sensitive to the first-order statistics of the input, i.e. the input mean, which may dominate the learning objective. Following our assumption that the goal of excitatory plasticity is to learn higher-order statistics, we hypothesize that neurons subtract the input mean, and, accordingly, we normalize inputs to zero mean in all our simulations. Short-term depression \citep{tsodyks_neural_1998} and threshold adaptation \citep{mensi_enhanced_2016} are candidate processes that might approximate input mean cancellation in cortical neurons. 

Eq.2-a is a plasticity rule combining nonlinear Hebbian potentiation with linear Hebbian depression. Here, nonlinear (or linear) refers to the quadratic (respectively linear) dependence upon the activity $y$ of the postsynaptic neuron. Importantly, the amplitude of the depression term is modulated by a metaplasticity function $h_y$ that tracks the squared rate of the postsynaptic activity, $\langle y^2 \rangle$, estimated in Eq.\ref{invrule}-b. We assume $\eta_h \gg \eta$ so that $h_y$ converges more rapidly than the weights. Note that we have recovered a variation of the BCM model \citep{intrator_objective_1992}, with a homeostatic factor $h_y = \langle y^2 \rangle$ instead of $h_y = \langle y \rangle^2$ in the original BCM model \citep{bienenstock_theory_1982}. The above arguments show that the generalized BCM models are part of a family of local learning models with the property of correlation-invariance.

\begin{figure}[!ht] 
\centering

\includegraphics[width=0.7\textwidth]{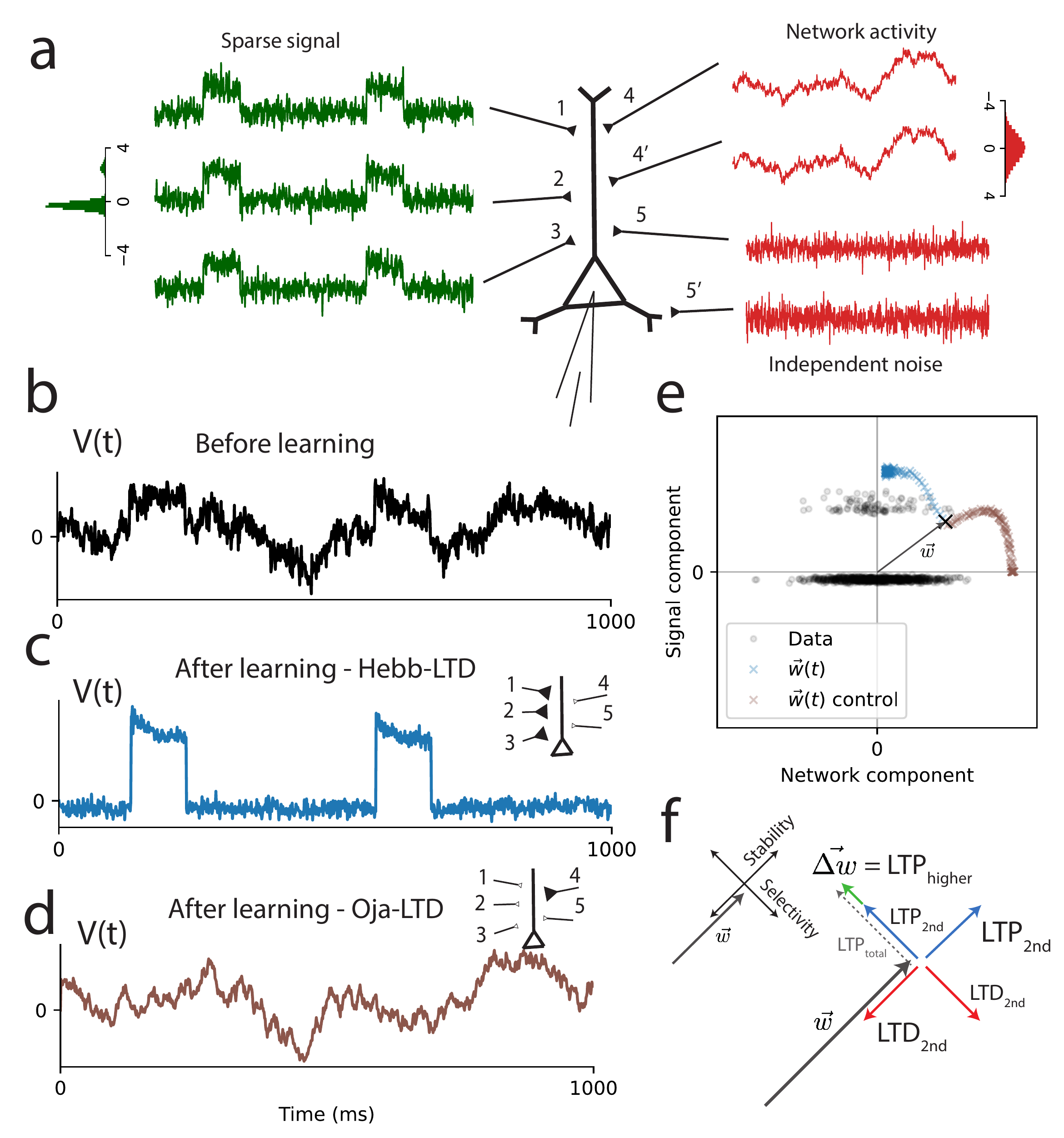}
\caption{\textbf{Learning sparse signals with correlation-invariance}. 
(\textbf{a}) Inputs belonging to three groups (20 inputs each): the sparse signal, with a non-Gaussian common component; the network activity, with a common Gaussian component, representing input from other brain areas; and independent background noise. Insets: histogram of amplitudes for sparse and network signals, with standard deviation $\sigma_{Network} = 1.2\ \sigma_{Sparse}$. (\textbf{b-d}) Membrane potential as a function of time, before learning (black), and after learning, for the correlation-invariant model (BCM rule, blue) and the model with Oja-like LTD (brown). Insets illustrate the synaptic strengths of each input group after learning.
(\textbf{e}) The learning dynamics of the weights (starting at the black X mark) projected to the Sparse and Network components. A subset of data samples is shown in grey. The correlation-invariant rule (blue) converges to the direction of sparsest activity, while the Oja-like rule (brown) converges to the direction of largest variance. This illustrates how the BCM model can perform Independent Component Analysis without a preprocessing step that decorrelates the inputs. (\textbf{f}) To illustrate the mechanism behind correlation-invariance, we decompose the weights $\bw$ into the stability and selectivity components. As the homeostatic mechanism balances LTP and LTD in the \emph{stability component}, the LTD term cancels the exact amount of second-order dependency of the LTP term. Since in the orthogonal direction (\emph{selectivity component}) the second-order components cancel as well, the net gradient $\Delta \bw$ (green) of the selectivity component depends only on the selectivity to higher-order statistics of the LTP term.
 }
\label{fig1}
\end{figure}

 We illustrate the effect of correlation-invariance in a neuron receiving inputs from three sources, including a group of 20 inputs with a common sparse signal of unit amplitude, another group of 20 inputs with a common high-amplitude Gaussian signal, and the third group with small background activity (Fig.\ref{fig1}-a). The correlation-invariant learning rule learns the sparse signal despite its low amplitude (Fig.\ref{fig1}-c,e). For comparison, we simulate a similar plasticity model, but with a heterosynaptic LTD mechanism adapted from the Oja learning rule \citep{oja_simplified_1982}, $\Delta \bw  = \eta \ (\bx \ y^2 - \bw y^2)$. Despite having a nonlinear LTP factor, this Oja-like model learns the high-amplitude Gaussian component (Fig.\ref{fig1}-d,e), illustrating that heterosynaptic LTD mechanisms provide stability, but not correlation-invariance.

\subsection*{Linear LTD enables correlation-invariance}

Numerous mechanisms have been proposed to account for the phenomenological properties of synaptic plasticity, but their specific properties and interactions are unclear \citep{pfister_optimal_2006,clopath_connectivity_2010,zenke_diverse_2015,keck_interactions_2017}. Our theory of correlation-invariant learning enables us to assign distinct functional roles to LTP, LTD, and homeostasis.

Critically, the LTD factor must be a depressive linear Hebbian factor, which is sensitive only to second-order correlations between pre- and postsynaptic neurons. In other words, the LTD factor must be proportional to $\bx y$ (and not to $\bx y^2$ or $\bx^2 y$). Let us recall the classic relationship between Hebbian learning and principal component analysis \citep{oja_simplified_1982}. The PCA algorithm maximizes the variance in the input, with an objective function $F(y) = \Exp{y^2}$, and can be implemented with a linear Hebbian learning rule, $\Delta \bw \propto \bx \ y$, with a positive proportionality constant. In contrast, in the correlation-invariant learning rule, the depression term is linear in pre- and postsynaptic activities, with a negative proportionality constant, $- \bx\ y$, which has the effect of removing the dependency on covariance from the learning rule,  which we may call an ``anti-PCA'' effect. Therefore the online learning procedure will learn the same features for raw inputs as it would for preprocessed inputs.

To have complete correlation-invariance, the LTD mechanism must cancel the correct amount of second-order dependency. We can show (Methods, Eqs.\ref{eq:h1}-\ref{eq:h2}) that this is exactly what happens when the homeostatic factor $h_y$ drives LTP and LTD to cancel each other in the direction of the weight vector. 
The component in the direction of the weight vector relates to the stability of the synaptic connections (i.e., the norm of the weight vector), and will be called 'stability direction' in the following. The orthogonal directions relate to feature
selectivity, determining which feature has been learned.
In Fig.\ref{fig1}-f, we give a geometric illustration for this mechanism in the 2-dimensional setting, decomposing the weights into the stability and selectivity components.
The key insight is that changes in the stability component only scale the inputs, affecting only second-order statistics, while not altering normalized higher-order statistics. When the norm of the synaptic weights is at its stable value, the LTD factor cancels the exact amount of the second-order dependency of the LTP factor in both components, leading to correlation-invariant learning. In contrast, Oja-like heterosynaptic LTD is proportional to the weight vector $\bw$ and does not act on the selectivity direction, leaving LTP selectivity dependent on second-order statistics.

\subsection*{Invariance to input amplitudes}

Cortical neurons receive inputs from presynaptic neurons with complex firing statistics. In general, synaptic plasticity models will fail to learn the expected features when different presynaptic neurons exhibit different scales of firing rate modulation, since classic Hebbian learning is sensitive to the activity level of presynaptic neurons \citep{oja_simplified_1982}. However, the correlation-invariant learning rule compensates for such differences, e.g. if the sparse signal arrives at the different synapses with different amplitudes (Fig.2-a,b). After learning, the synaptic weights are inversely proportional to the input amplitudes (Fig.2-c), resulting in each input having the same contribution to the total input current. We can compare the plasticity model with an optimal linear decoder, trained with linear regression to output the sparse latent feature. We see that the correlation-invariant model achieves almost the optimal recovery of the latent signal (Fig.2-e). 

This invariance may be relevant for neurons with a large dendrite. For instance, the effect of input spikes on the somatic membrane potential is scaled down by dendritic attenuation, which varies with the distance from the synapse to the soma. It has been observed that synaptic strengths compensate for dendritic attenuation, and distal synapses have the same level of depolarization as proximal ones \citep{magee_somatic_2000}. This is expected in the presence of a correlation-invariant plasticity mechanism, which self-organizes the synaptic weights to compensate for linear disparities between synaptic inputs. 
Importantly, and in contrast with earlier work \cite{chance_gain_2002}, our synaptic plasticity rule compensates for the difference in mean drive while staying sensitive to sparse features in the input.

\begin{figure}[!ht] 
\centering

\includegraphics[width=
\textwidth]{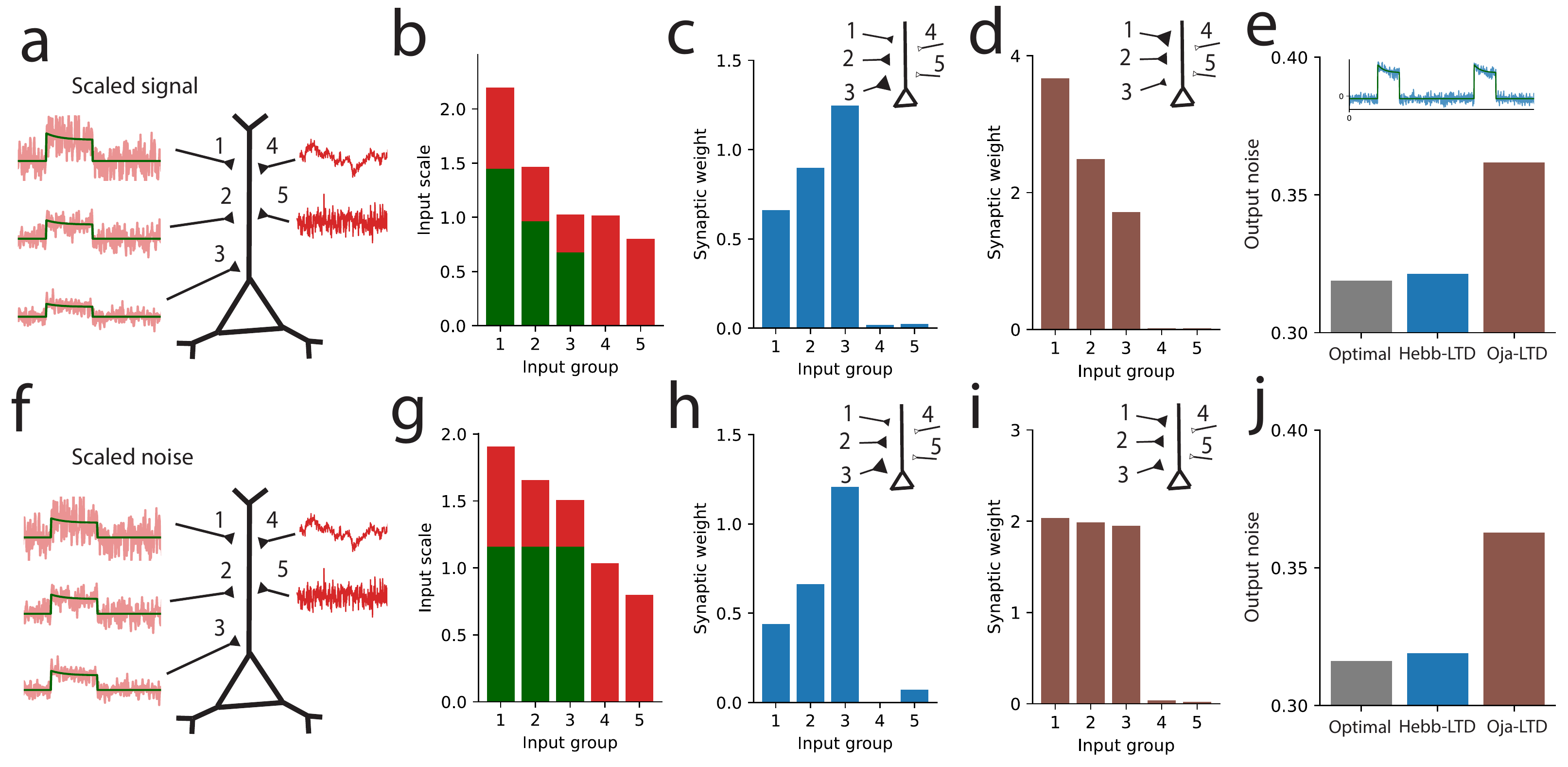}

\caption{\textbf{Optimal decoding under variable input scaling and noise}. (\textbf{a}) Sparse inputs with different amplitude levels. (\textbf{b}) The sparse component group is divided into three subgroups with different standard deviations for their signal (green) and noise (red) levels, with the same signal-to-noise ratios. (\textbf{c}) The correlation-invariant rule learns weights that compensate for the input scaling, with final weights inversely proportional to the input signal amplitude. (\textbf{d}) The model with Oja-like LTD learns weights that are proportional to the input amplitudes. (\textbf{e}) Inset: Output activity after learning with the correlation-invariant rule. Main graph: Remaining output noise estimated from the deviation between the (noiseless) sparse signal and the optimal linear decoder trained on the 5 groups of input channels (optimal, left), the neuronal output after learning with the correlation-invariant rule (Hebb-LTD, middle/blue), or with Oja-LTD (right/brown).
(\textbf{f}) Sparse inputs with different noise levels. (\textbf{g}) The sparse component group is divided into three subgroups with different noise levels (red), but the same signal amplitude (green). (\textbf{h}) The correlation-invariant rule decodes the signal, learning synaptic weights proportional to the input signal-to-noise ratios. (\textbf{i}) For comparison, the rule with Oja-like LTD learns weights proportionally to input signal amplitude. (\textbf{j}) As above, the correlation-invariant rule converges to a decoder almost as efficient as the optimal linear decoder.
 }
\label{fig-components}
\end{figure}

\subsection*{Optimal decoding from noisy inputs}

When performing inference about a sensory variable, the brain integrates information from multiple unreliable sources, weighing them according to their reliability \citep{graf_decoding_2011, fetsch_neural_2012}. To learn such an efficient decoder, neural circuits must be able to adapt incoming synapses according to the information conveyed by each input, searching for the input combination with the highest signal-to-noise ratio. Conveniently, when decoding sparse latent variables, the direction with the highest signal-to-noise ratio will also be the direction with the sparsest distribution. Thus we can use our sparse learning objective to recover the most informative direction, using the correlation-invariant learning rule to learn an efficient decoder.

We simulated a neuron for which the inputs have variable signal-to-noise ratios (Fig.\ref{fig-components}-f,g). The correlation-invariant learning rule develops weights proportional to the input signal-to-noise ratio, giving more importance to more informative inputs, leading to an output noise level close to the optimal linear decoder (Fig.\ref{fig-components}-h,j). Importantly, the plasticity rule does not simply select the one input synapse that has the highest signal-to-noise ratio but selects all input synapses that carry the signal, albeit with different importance weights. On the other hand, the learning rule with heterosynaptic LTD learns weights proportional to the input signal amplitude, with little sensitivity to signal-to-noise levels (Fig.\ref{fig-components}-i,j). 

These results suggest that correlation-invariance could be a fundamental learning mechanism underlying near-optimal decoding from sensory information and multi-sensory integration, as seen in experiments \citep{graf_decoding_2011, fetsch_neural_2012}. In comparison with related models based on maximal information transmission, such as independent component analysis \citep{bell_independent_1997}, the correlation-invariant model requires minimal assumptions on the input distribution. A single plasticity rule learns an efficient decoder for different input scales, noise levels and sparse latent distributions.

\subsection*{Learning sparse population codes from correlated inputs}

While so far we have considered the learning properties of single neurons, sensory networks contain populations of neurons, with different tuning curves for each neuron in the population. with each neuron in the population representing different parts of the latent space, illustrated by the tuning curves of the population. For a given ensemble of inputs, uncovering an efficient population code can be challenging. 
For instance, for optimal information transmission, it is impotant to distinguish between relevant variables that characterize the signal and unrelated input variability that can be seen as noise \citep{pouget_inference_2003}. We assume that the signal is characterized by a sparse, non-Gaussian distribution. We show that under this assumption the correlation-invariant model is able to learn efficient population codes.

\begin{figure}[!ht] 
\centering

\includegraphics[width=\textwidth]{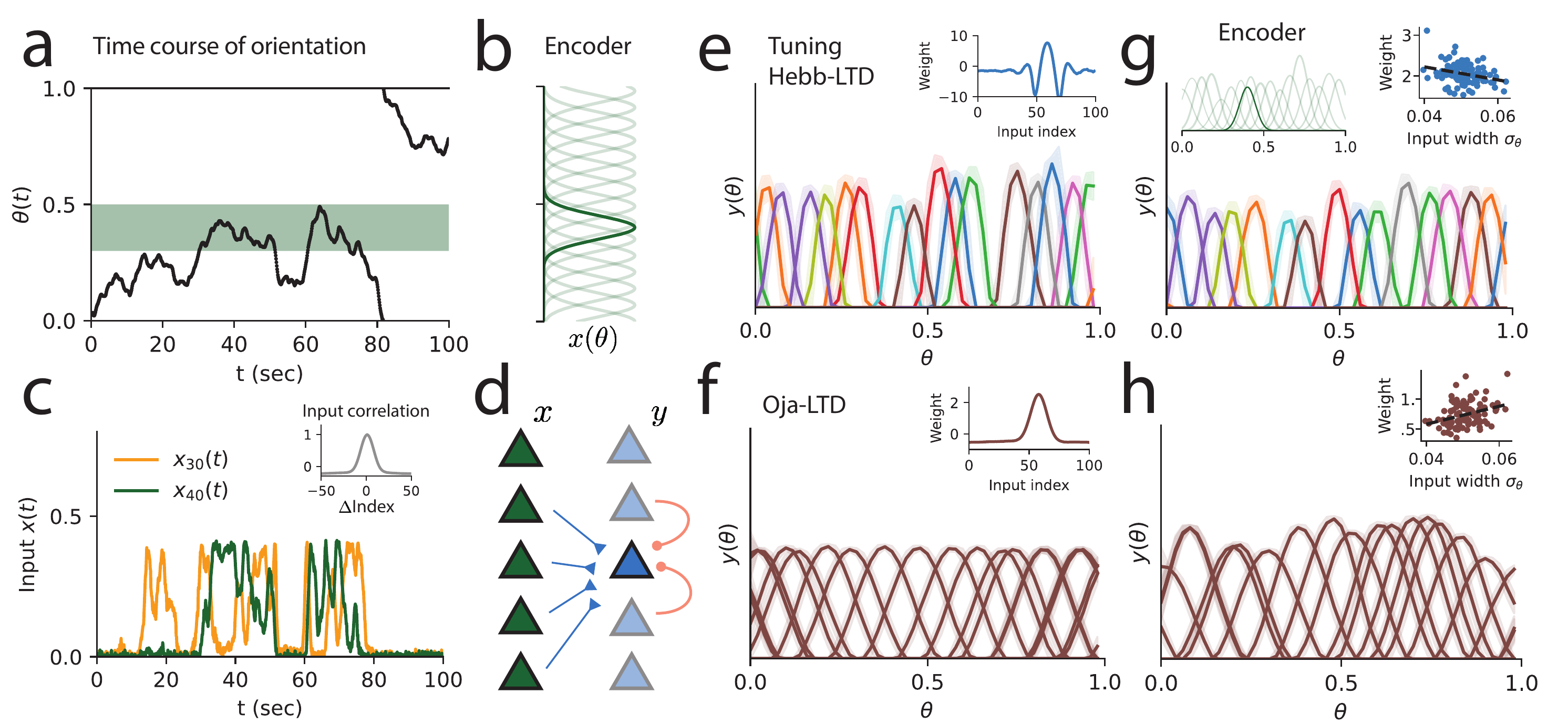}
\caption{\textbf{Correlation-invariant dictionary learning in a population coding network}. (\textbf{a}) A circular continuous latent variable follows a random walk with values between $[0,1]$. (\textbf{b}) The input population $x$ encodes the latent variable with N=100 Gaussian tuning curves $x(\theta(t))$. (\textbf{c}) The activity of two input neurons over time. Nearby inputs show positive correlations, following their overlap in tuning (inset, grey). (\textbf{d}) Network diagram, with the input population (green) projecting synapses to a decoding population (blue). The synapses change according to the synaptic plasticity model and can take positive or negative values. Recurrent inhibition is included between all neurons (orange). (\textbf{e}) The correlation-invariant model learns a dictionary of Mexican hat-like synaptic weights (inset, blue), inverting the input correlation profile, with tuning curves tiling the latent space with small overlaps between response profiles (coloured, variability in light shade). The population tuning curves are sharper (mean width at half maximum $\sigma_{\theta} = 0.07$) than the tuning of input neurons ($\sigma_{\theta} = 0.11$). (\textbf{f}) With Oja-LTD, neurons in the population learn synaptic weights (inset) following the input correlations, with wider tuning curves ($\sigma_{\theta} = 0.17$) than that of input neurons. (\textbf{g}) We simulate a new input population with heterogeneous tuning curves, with variation in width, amplitude and noise levels (inset, green). The correlation-invariant model learns again a sparse dictionary, optimizing for the sparsest, lowest-noise representation. Tuning curves are sharper ($\sigma_{\theta} = 0.08$) than the input tuning ($\sigma_{\theta} = 0.11$), with higher selectivity for sharper input neurons (inset, correlation between input tuning width $\sigma_{\theta}$ and synaptic weight magnitudes: $\rho_{\sigma w} = -0.28$). (\textbf{h}) With Oja-LTD, the population dictionary follows the input variance and correlation profile, learning wide tuning curves ($\sigma_{\theta} = 0.14$), with higher selectivity to wider tuned input neurons (inset, $\rho_{\sigma w}=+0.34$).
 }
\label{fig3}
\end{figure}

We consider a line stimulus (or Gabor patch stimulus) that changes its orientation slowly over time. The stimulus is encoded by a noisy input population, with input tuning curves tiling the space of orientation angles (Fig. \ref{fig3}-a,b). As each input neuron is selective to only a part of the input space, they show sparse activity, with the overlap of the tuning curves generating positive input correlations between neighbouring neurons (Fig. \ref{fig3}-c). 

We extend our single neuron model to a population of output neurons, with synapses from input to output population following the correlation-invariant plasticity rule (Fig. \ref{fig3}-d). In cortical networks, recurrent inhibition is thought to decorrelate excitatory neurons, thereby allowing them to learn different features \citep{foldiak_forming_1990,vogels_inhibitory_2011,boerlin_predictive_2013}. We thus include inhibitory recurrent connections between output neurons, which we consider as a simplified effective description of the local excitatory-inhibitory network \citep{wong_recurrent_2006}. Recurrent connections change with a covariance-based plasticity rule \citep{vogels_inhibitory_2011}. To avoid dynamic instabilities due to concurrent excitatory and inhibitory plasticity, we include multiplicative weight decay in both \citep{turrigiano_homeostatic_2004}.

After learning, output neurons developed Mexican hat-like synaptic weight profiles, which have the effect of cancelling input correlations, leading to a population code tiling the space of line orientations with tuning curves sharper than those of inputs in the input layer (Fig. \ref{fig3}-e, mean tuning width $\sigma_\theta = 0.07$; input tuning width $\sigma_\theta = 0.11$). Following the same learning principles as in the single neuron case, the population code developed through learning can be interpreted as an efficient code with minimal redundancy. In comparison, the Oja-like learning rule learns wider tuning curves, which follow input directions of large variance, dominated by the input correlations (Fig. \ref{fig3}-f, $\sigma_\theta = 0.17$). 

Under more realistic conditions, sensory populations must decode information from neurons with diverse tuning properties. As we expect the correlation-invariant rule to be invariant to such input properties, we test the plasticity model in the presence of input heterogeneities. We simulated input tuning curves of variable widths, amplitudes and noise levels. As seen in Fig. \ref{fig3}-g, the correlation-invariant model learns a population code with similar properties ($\sigma_\theta = 0.08$) as for homogeneous input tuning, with higher selectivity for more precise input neurons. On the other hand, a model without correlation-invariance learns wider tuning curves ($\sigma_\theta = 0.14$), dependent on the input tuning profiles. In particular, neurons develop more selectivity for input neurons with wider tuning, disregarding their precision (Fig. \ref{fig3}-h).

\subsection*{Spiking model of sensory development with correlated inputs}

\begin{figure}[h] 
\centering

\includegraphics[width=\textwidth]{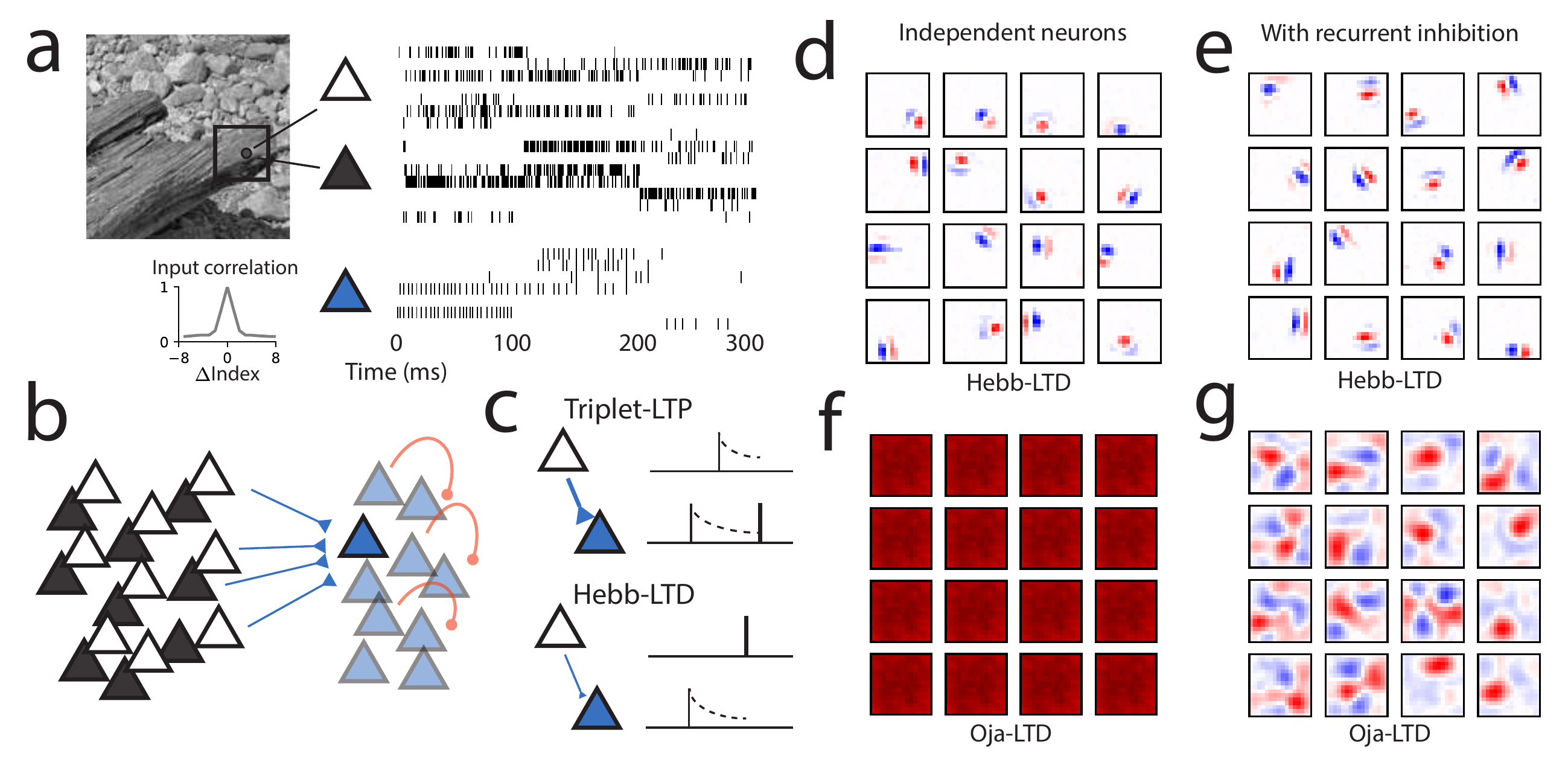}
\caption{\textbf{Correlation-invariant learning with triplet STDP facilitates receptive field development in a spiking network}. 
(\textbf{a}) Inputs were sampled from 16x16 patches of natural images (left), encoded as ON/OFF population with Poisson spiking rates (right), representing visual input projections. The input has high pair-wise correlations for nearby pixels, and positive correlations over the whole patch (inset, grey)  (\textbf{b}) Spiking network model, with inputs projecting feed-forward excitatory weights to a population of 64 spiking neurons and recurrent inhibition. (\textbf{c}) Excitatory weights are modified through triplet STDP, including the LTD mechanism linear on pre-post spiking correlations. (\textbf{d}) Showing correlation-invariance, spiking neurons with triplet STDP learn localized receptive fields even in the absence of lateral inhibition, despite input correlations. (\textbf{e}) With recurrent inhibition included, neurons still learn similar receptive fields. (\textbf{f}) The spiking version of the Oja-like model, with heterosynaptic LTD, learns non-local input projections, due to sensitivity to input correlations. (\textbf{g}) In the absence of correlation-invariance, lateral inhibition can promote somewhat more localized receptive fields, though still sensitive to the input correlation profile.
 }
\label{fig-multiple}
\end{figure}

TThe relevance of a plasticity model comes from both the biological plausibility of the plasticity rule and from emerging functionality when embedded in plausible networks of spiking neurons. The correlation-invariant learning rule has a solid foundation in plasticity rules extracted from experimental data on cortical excitatory synapses. Cortical development is driven by voltage-dependent and spike-timing-dependent-plasticity (STDP), with synaptic changes depending on the relative timing of pre and post-synaptic spikes \citep{markram_history_2011}. In particular, plasticity in excitatory synapses is well modelled by the voltage-based Clopath model \citep{clopath_connectivity_2010} or the triplet STDP model \citep{pfister_triplets_2006}, in which LTP depends on one pre- and two post-synaptic spikes, and LTD on single pre- and post-synaptic spikes (Fig. \ref{fig-multiple}-c). Considering a Poisson firing regime, and a homeostatic mechanism, the triplet model can, under rather general assumptions, be reduced to the rate model we have considered so far, $\Delta \bw = \eta \ ( \bx \ y^2 - h_y \ \bx \ y)$ \citep{clopath_connectivity_2010, gjorgjieva_triplet_2011}. 
From this relation, we might expect a spiking model of sensory development with triplet STDP to show correlation-invariance. Relative to rate models, spiking models are notoriously challenging to train, with added difficulty including spiking variability and spike-spike correlations. Additionally, it constrains the input representation to be non-negative, changing how sensory information is processed.  

We implemented a spiking network for sparse population coding, modelling V1 receptive field development from natural images (Fig. \ref{fig-multiple}-a,b). It is a classic example where neurons develop selectivity to specific properties in their input, with the network creating a dictionary of localized orientation-selective features, which are the sparse features of natural images \citep{field_what_1994, olshausen_emergence_1996}. While many models have been shown to develop such sparse codes, they generally assume that the input has been preprocessed to be decorrelated and normalized \citep{brito_nonlinear_2016}. Though the retinal pathway is known to partially decorrelate the visual stimuli, the input to cortical neurons still maintains some degree of correlation. In the presence of spatial correlations, other models relied on recurrent inhibition, which diversifies the features learned by the network. In situations where single neurons or small  networks  learn the principal components (non-localized spatial Fourier filters) of the input
images,  features of sparse coding appeared only if recurrent inhibition was strong and the network large enough \citep{olshausen_emergence_1996, zylberberg_sparse_2011}. Motivated by the correlation-invariant theory, we wanted to test whether our plasticity rule can learn localized filters directly from image data without pre-whitening.

We considered an input dataset of natural image patches, encoded into ON and OFF spiking inputs, showing positive input correlations for neighbouring pixels (Fig. \ref{fig-multiple}-a). Similarly to the rate model, we implement triplet STDP on input to output connections, output neurons modelled as leaky integrate-and-fire, and recurrent inhibition with inhibitory plasticity \citep{vogels_inhibitory_2011} (Fig. \ref{fig-multiple}-b, see Methods).

To probe if learning was possible with single output neurons, we first ran the model without lateral inhibition. After learning, neurons developed localized receptive fields, composed of ON and OFF parts, similar to what is observed in V1 (Fig.\ref{fig-multiple}-d), showing that the model can develop sparse features even without lateral inhibition. Even though each neuron has identical inputs, we see a diversity of receptive fields due to random initial conditions of the synaptic weights. When lateral inhibition was included, the model learned a similar dictionary of localized filters (Fig.\ref{fig-multiple}-e). While lateral inhibition was not necessary in this setting for learning localized filters, it ensures the diversity of receptive fields \citep{olshausen_emergence_1996, brito_nonlinear_2016}.
 
 We compared our results with those of an Oja-type STDP model where LTD is implemented as  heterosynaptic plasticity and found that in this case all neurons learned a non-localized receptive field covering the whole patch, as is expected for a principal component of the input (Fig.\ref{fig-multiple}-f). Only when lateral inhibition was included, did ON/OFF receptive fields appear, though not completely localized  (Fig.\ref{fig-multiple}-g). It demonstrates that without correlation-invariance, the spiking model is sensitive to input correlations, and requires lateral inhibition to enforce the tiling of the input space into sparse tuning curves. These results indicate that while correlation-invariance can be sufficient for learning sparse tuning curves, lateral inhibition can produce similar effects, with both mechanisms potentially at work in parallel in cortical circuits. In summary, by adding robustness to input noise, to input scaling and to second-order input correlations, the spike-based version of the correlation-invariant rule supports spiking network to develop sensory representations with a diversity of localized receptive fields.





\section*{Discussion}

We have presented correlation-invariance as a critical property of cortical synaptic plasticity. Correlation-invariance is derived from the normative perspective of feature learning, in which cortical neurons develop responses to sparse latent features \citep{foldiak_forming_1990, olshausen_emergence_1996}. 
Though many models that develop sparse features have been proposed, they have in general sidelined the problem of input correlations by artificially pre-whitening inputs. We have shown that we can extend the contexts under which plastic neurons can learn useful representations by considering a biologically plausible plasticity model that discounts second-order statistics.

Correlation-invariance stands in contrast to the original Hebbian learning perspective, grounded on learning by association \citep{hebb_organisation_1952, oja_simplified_1982}. Instead, correlation-invariant models discount linear correlations, learning only higher-order correlations. The critical mechanism is a linear LTD factor, which has been observed in cortical excitatory synapses \citep{pfister_triplets_2006}, for which our results suggest a functional explanation. Our theory extends our previous understanding of Hebbian mechanisms and may aid the development of more complex representation learning models.

\subsection*{A unifying theory for models of synaptic plasticity}

Hebbian models such as BCM and Oja learning rules are decades old, and many studies have investigated their functional properties, concerning their stability, feature selectivity and receptive field development \citep{oja_learning_1991,blais_receptive_1998,cooper_bcm_2012}. Nevertheless, the functional difference between Oja's heterosynaptic depression and BCM's anti-Hebbian depression factor has remained unclear. Our analysis shows that linear LTD allows for correlation-invariance, while Oja's heterosynaptic depression only acts on the stability component. These studies can now be unified in a theoretical framework, with their empirical observations predicted by the theory. 

We have also uncovered an interesting relation between BCM, ICA and sparse coding, classic models of early sensory development. ICA and sparse coding start from similar normative assumptions, with inputs as mixtures of latent sparse features \citep{olshausen_sparse_1997}. Our normalized objective function can be seen as an alternative formulation of sparse coding, for single neurons \citep{brito_nonlinear_2016}. Though the BCM model was first proposed as a stable version of Hebbian learning, we have shown that it links naturally to a normative formulation of sparse feature learning, with each of its elements seemingly designed for this task. We believe our theory provides a systematic basis for the analysis and development of Hebbian plasticity models.

Though our theory is based on a single-neuron objective, our network simulations demonstrate that correlation-invariant learning is compatible with learning network representations. It is essential to investigate how the theory of correlation-invariance might be integrated with related normative models for learning sparse, efficient representations \citep{zylberberg_sparse_2011,boerlin_predictive_2013}.

\subsection*{Correlation-invariance in cortical neurons}

The correlation-invariant learning rule has a precise correspondence to phenomenological models of spike-timing-dependent plasticity, including the triplet and voltage-dependent STDP models, which reduce to a quadratic postsynpatic factor for LTP and a linear postsynaptic factor for LTD \citep{pfister_triplets_2006, clopath_connectivity_2010}. In particular, our theory suggests that pyramidal neurons should include synaptic LTD mechanisms linear in both pre and post-synaptic activities, in agreement with experimental findings \citep{pfister_triplets_2006}. Pairing experiments under Poisson firing times of pre and post-synaptic neurons would be valuable to investigate to what extent these properties hold \citep{froemke_spike-timing-dependent_2002}. 

Since traditional metaplasticity experiments have searched on slow time scales \citep{turrigiano_homeostatic_2004}, it is unclear whether a rate detector exists that is fast enough to fulfil the function of the homeostatic factor $h_y$ \citep{zenke_synaptic_2013}. 
Nevertheless, stability may also be achieved through other mechanisms, such as heterosynaptic plasticity, though in this case, correlation-invariance will be partial and dependent on input statistics. In this case, there will be a compromise between learning higher-order and second-order statistics. Plasticity sensitivity to second-order statistics might be useful for other tasks, such as learning associative memories \citep{zenke_diverse_2015}.

Some findings on synaptic weight distribution provide evidence that cortical synapses self-organize with correlation-invariance. It has been observed that distal synapses are relatively up-regulated compared to proximal, and have in general somatic effects at the same order of magnitude as proximal connections \citep{magee_somatic_2000}. Experiments on how synaptic profiles depend on input firing rates and correlations would be ideal to probe to which extent correlation-invariance might be at work in cortical circuits.

\subsection*{Learning efficient population codes under diverse conditions}

 Experimental evidence indicates that primates can combine unreliable sensory information as would a near-optimal decoder \citep{graf_decoding_2011, fetsch_neural_2012}. Normative population coding models approach this task by defining what each neuron represents about stimuli, for instance, the log-likelihood \citep{jazayeri_optimal_2006} or a probability distribution \citep{ma_bayesian_2006}, from which a decoder can be designed. Such a design is difficult to learn with local rules, especially if the input can have unknown levels of reliability and correlations \citep{pouget_inference_2003}. 
 
  Instead, the correlation-invariant sparse objective operates at the algorithmic level, with minimal assumptions about how the input represents the latent variable. By assuming sparse latent variables, the objective becomes equivalent to maximizing information transmission, enabling the development of population codes with sharp tuning and low noise. How these sparsity-based models relate to other normative population coding models is an important topic for further investigation.

\subsection*{Search for biological learning algorithms}

Representation learning is a difficult task and it is puzzling how the brain is capable of developing, maintaining and adapting a complex model of the external world. Only recently have artificial learning models been able to learn with very large, complex networks, but with methods that are not easily mapped to biological mechanisms \citep{lecun_deep_2015,lillicrap_backpropagation_2020}.

In the absence of supervising signals, unsupervised Hebbian plasticity provides the framework for learning a representation and may underlie how the cortex learns through local information \citep{illing_local_2021, halvagal_combination_2022, brucklacher_local_2022}. Reinforcement learning is another central paradigm for understanding biological learning, believed to have a biological instantiation in neuromodulators and reward modulated plasticity. Indeed there is evidence in favour of the influence of reward-based learning on input representations and receptive fields in sensory cortices \citep{shuler_reward_2006, poort_learning_2015}. It is an active field of research on how neuromodulators interact with Hebbian mechanisms \citep{fremaux_functional_2010,gerstner_eligibility_2018,aljadeff_cortical_2019}. It would interesting to see how theories of sparse feature learning and correlation-invariance might be integrated with reinforcement learning objectives.  Correlation-invariance extends the theory and function of Hebbian plasticity and might be an additional building block for models and theories of biological learning \citep{marblestone_toward_2016}. 







\section*{Methods}
\label{sec:methods}

\subsection*{Linear invariance of the normalized objective function}

We consider the normalized projection pursuit objective, of the form 
\begin{equation}
\bw^* = \text{argmax}_{\bw} \ \Exp{F \left(\frac{\bw^T \bx}{\sigma}\right)}
\label{normobj}
\end{equation}
with $\sigma = \sqrt{\Exp{(\bw^T \bx)^2}}$. Let $\mathbf{M}$ be a  transformation matrix for $\bx$ that makes it decorrelated:
\begin{align}
\tilde{\bx} = \mathbf{M} \bx \implies \Exp{\tilde{\bx} \tilde{\bx}^{T}} = I
\label{decorr}
\end{align}
The transformation $M$ is called \emph{whitening} \citep{hyvarinen_fast_1999}. For instance, we can construct 
$\mathbf{M} = \mathbf{R}\mathbf{D}^{-1/2}\mathbf{R}^{T}$, where $D$ is a diagonal matrix and $\langle\mathbf{x}\mathbf{x}^{T}\rangle=\mathbf{R}\mathbf{D}\mathbf{R}^{T}$
is the eigenvalue decomposition of the input correlation matrix.  Using that $\bx = \mathbf{M^{-1}} \tilde{\bx}$ and defining $\tilde{\bw} = \mathbf{M}^{-T} \bw$, we have
\begin{align}
\Exp{F \left(\frac{\bw^T \bx}{\sigma}\right)} 
= & \Exp{F \left(\frac{\bw^T \mathbf{M^{-1}} \tilde{\bx}}{\sqrt{\Exp{(\bw^T \mathbf{M^{-1}} \tilde{\bx})^2}}}\right)} \\
= & \Exp{F \left(\frac{\tilde{\bw}^{T} \tilde{\bx}}{\sqrt{\Exp{(\tilde{\bw}^{T} \tilde{\bx})^2}}}\right)} \\
= & \Exp{F \left(\frac{\tilde{\bw}^{T}}{|\tilde{\bw}|} \tilde{\bx}\right)}
\end{align}
where we used Eq.\ref{decorr} to simplify the denominator: $\Exp{(\tilde{\bw}^{T} \tilde{\bx})^2} = \Exp{\tilde{\bw}^{T} \tilde{\bx} \tilde{\bx}^{T} \tilde{\bw}} = \tilde{\bw}^{T} \Exp{\tilde{\bx} \tilde{\bx}^{T}} \tilde{\bw} = \tilde{\bw}^{T} \tilde{\bw} = |\tilde{\bw}|^2$. Thus the normalized objective function can be mapped to a standard objective function, with normalized weights and whitened inputs $\tilde{\bx}$,
\begin{equation}
\tilde{\bw}^* = \text{argmax}_{\tilde{\bw}, |\tilde{\bw}| = 1} \ \Exp{F \left(\tilde{\bw}^{T} \tilde{\bx}\right)}
\label{eq:wpp}
\end{equation}
with an optimum in the original input space given by $\bw^* = \mathbf{M}^T \tilde{\bw}^*$. 

Analogously, given any linear transformation of the input, $\bx' = \mathbf{R} \bx$, for an invertible matrix $\mathbf{R}$, we may map the normalized projection pursuit to the whitened projection pursuit of Eq.\ref{eq:wpp}, with the optima given by $\bw'^* = \mathbf{R}^{-T} \mathbf{M}^T \tilde{\bw}^*$. Hence, the normalized objective function of Eq.\ref{normobj} is invariant to linear transformations of the input.


\subsection*{A correlation-invariant rule with arbitrary norm $|\bw|$}

We consider $F(a) = a^3$  with $a = \bw^T \bx / \sqrt{\langle ( \bw^T \bx  )^2 \rangle}$ and search for the optimal weight vector
\begin{equation}
  \bw^* = \text{argmax}_{\bw} \ \Exp{ \left(\frac{y}{\sigma_y} \right)^3 }
\end{equation}
assuming that the neuron has a rectified linear activation function $y = (\bw^T \bx)_+$ and where $\sigma_y = \sqrt{\langle y^2 \rangle}$.

Proceeding with gradient ascent on $\bw$, we have
\begin{align}
\frac{\partial \Exp{F}}{\partial \bw} &= \frac{\partial }{\partial \bw}  \Exp{ \left(\frac{y}{\sigma_y}\right)^3 } \\
&= 3 \Exp{ \left(\frac{y}{\sigma_y}\right)^2 
\left(\sigma_y^{-1}  \frac{\partial y}{\partial \bw}  + 
y \frac{\partial \sigma_y^{-1}}{\partial \bw}  \right) } \\
&= \frac{3}{\sigma_y^2} \Exp{ y^2 \left(\frac{1}{\sigma_y} \frac{\partial y}{\partial \bw}  -
\frac{y}{\sigma_y^2} \frac{\partial \sigma_y}{\partial \bw}\right) } 
\end{align}

We now use that the neuron has a rectified linear activation function, so that $\frac{\partial y}{\partial \bw} = \bx_+$ and $\frac{\partial \sigma_y}{\partial \bw} = \frac{\partial \sqrt{\langle y^2 \rangle}}{\partial \bw} = \langle y \bx_+ \rangle / \sigma_y$, where we define $\bx_+ = \bx \ \textbf{I}_{y>0}$ as the input for samples in which $y \ge 0$. 
Since the output of the neuron is always non-negative, we have $y \ge 0$ for all $\bx$ so that we have $x_+ y = x y$ and $x_+ y^2$ = $x y^2$. This yields
\begin{align}
\frac{\partial \Exp{F}}{\partial \bw}  &= \frac{3}{\sigma_y^2} \Exp{ \left(\frac{\bx_+ y^2}{\sigma_y} -
\frac{y^3}{\sigma_y^3} \langle \bx_+ y \rangle \right) } \\
&= \frac{3}{\sigma_y^3} \left( \langle \bx y^2 \rangle -
\frac{\langle y^3 \rangle}{\langle y^2 \rangle} \langle \bx y \rangle \right)
\end{align}

To derive an online learning rule, we consider a separation of time scales and assume that the estimation of $\sigma_y$ and $\frac{\langle y^3 \rangle}{\langle y^2 \rangle}$ is performed at a faster time scale than the other factors, which allows us to consider them as constants. We derive a stochastic gradient ascent learning dynamics by removing the estimation over the whole dataset,
\begin{equation}
\Delta \bw \propto \bx \ y^2 - h_y \ \bx \ y 
\end{equation}
We refer to the specific choice $h_y^* = \frac{\langle y^3 \rangle}{\langle y^2 \rangle}$ as the \emph{balancing homeostatic factor}. We claim that the balancing homeostatic factor leaves the learning rule at an \emph{indifferent stability} in the direction of the weights, leaving the norm fluctuating freely. 
We can check this property by showing that the gradient in the direction of the synaptic connections is zero, 
\begin{equation}
\langle \bw^T \Delta \bw \rangle \propto \langle y^3 \rangle - h_y^* \langle y^2 \rangle = 0
\end{equation}
It is a consequence of using an objective function that is invariant to the norm of the weight vector.

\subsection*{A family of correlation-invariant learning rules with stable weights}

While the top-down derivation of the correlation-invariant learning rule leads to a specific balancing homeostatic factor $h_y^* = \frac{\langle y^3 \rangle}{\langle y^2 \rangle}$, it is not a stable learning rule, as the norm will vary freely. Instead we can consider factors that are stable, such as $h_y = \langle y^2 \rangle$. In fact any supralinear factor $h_y = \langle y^r \rangle$, with $r > 1$, will lead to stable dynamics \citep{intrator_objective_1992}.
We claim that the family of stable plasticity rules with these alternative homeostatic factors will, after convergence, optimize the same objective function as
the learning rule derived in the previous paragraph. 
To demonstrate this, we calculate the homeostatic factor once the norm has converged to a stable value. Under this assumption, the gradient in the direction of the weights $\bw$ is zero, 
\begin{equation}
\langle \bw^T \Delta \bw \rangle \propto \langle y^3 \rangle - h_y \langle y^2 \rangle = 0 \implies h_y = \langle y^3 \rangle / \langle y^2 \rangle = h_y^*(y) 
\label{eq:h1}
\end{equation}
which implies that when the norm has converged to a stable value during the learning process, the stabilizing homeostatic factor $h_y$ will have the same value as the balancing homeostatic factor $h^*_y$ for the same weights, and consequently will have the same correlation-invariant properties.

We can also calculate analytically the norm the weights will have during the learning process. For $h_y = \langle y^2 \rangle$, we have
\begin{align}
  h_y = \langle y^3 \rangle / \langle y^2 \rangle &\iff \langle y^2 \rangle = \langle y^3 \rangle / \langle y^2 \rangle  \\
   &\iff |\bw|^2 \langle x_w^2 \rangle = |\bw| \langle x_w^3 \rangle / \langle x_w^2 \rangle \\
  &\iff |\bw| =  \langle x_w^3 \rangle / \langle x_w^2 \rangle^2 \label{eq:h2}
\end{align}
where $x_w = (\bw^T \bx)_+ / |\bw|$ is the rectified projection of the input $\bx$ on the normalized direction $\bw/|\bw|$.

Importantly, the norm of the weight vector does not converge to a predefined value, e.g. $|\bw| = 1$ as in the simple model of Eq.\ref{nhl} or in the Oja rule \cite{oja_simplified_1982}, but has a final value that depends on the input statistics.

\subsection*{Simulations}

For the single neuron simulations, we generated three input groups of 20 neurons each. The sparse signal had ON states with a duration of 100ms, with interstimulus intervals following an exponential distribution (time scale $\tau_1 = 1000ms$), and added independent Gaussian noise to each neuron. The network signal followed an Ornstein-Uhlenbeck process (time scale $\tau_2 = 200ms$), and added independent Gaussian noise. The third group of inputs was generated as independent Gaussian noise. All inputs were mean subtracted. For Fig. 1, input standard deviations of each group were $\sigma_1 = 1.$, $\sigma_2 = 1.2$, $\sigma_3 = 2.2$, respectively. For Fig. 2-a,b, the sparse signal inputs were subdivided in three groups with different amplitudes, $\sigma_{11} = 1.5$, $\sigma_{12} = 1.$, $\sigma_{13} = 0.7$. For Fig. 2-f,g, the sparse signal inputs were subdivided in three groups with different independent noise amplitudes, $\sigma^{n}_{11} = 1.5$, $\sigma^{n}_{12} = 1.$, $\sigma^{n}_{13} = 0.7$. 

The homeostatic factor $h_y = \langle y^2 \rangle$ was estimated as a moving average of $y^2$ with time scale of $\tau_h = 200$ samples: $h_{t} = h_{t-1} \ (1 - 1/\tau_h) - y^2_t / \tau_h$. All simulations generated $10^6$ data samples and ran the learning model for $10^6$ time steps. We implemented stochastic gradient descent updates using the Adam optimizer with learning rate $\eta = 0.003$, mini-batches with 100 random samples, and random initial weights with a Gaussian distribution of mean zero and unit variance. 

For the population coding simulations, we generated the latent variable from a random walk, smoothed with an exponential filter (time scale $\tau_3 = 100ms$), with circular values, by clipping to $[0,1]$. We generated 100 inputs, with evenly spaced Gaussian tuning curves, with $0.05$ width, including additive independent Gaussian noise to the input activities ($\sigma = 0.01$). For generating heterogeneous tuning curves, we scaled the noise, width and amplitude of each tuning curve by independent log-normal random variables, with zero mean and $\sigma=0.2$. The population network included 16 output neurons. We included all-to-all inhibitory recurrent connections $\bw^{rec}_ {ij}$ from neuron $j$ to neuron $i$, without self-connections. Each neuron had activation $y_j = (\bw^T \bx + \bw_{rec}^T \mathbf{y})_+$, with inhibitory plasticity $\Delta \bw^{rec}_{ij} = -\eta^{rec} (y_i (y_j - \theta) - \lambda^{rec} \bw_{ij}^{rec})$, clipped to negative values only, with $\lambda^{rec} = 1.0$, $\theta = 1.$, $\eta^{rec} = 0.03$. To maintain network stability, we also added weight decay to the feedforward plasticity model, $\Delta \bw_t = \eta (\bx_t y_t^2 - h_y \bx_t y_t - \lambda \bw)$, with $\lambda = 0.001$. For each input sample, we ran the recurrent dynamics for 10 time steps. 

For the spiking network, we generated 16x16 image patches, sampled from black and white natural images \citep{olshausen_emergence_1996}, divided into ON and OFF cells, totalling 512 input neurons. Input spike trains were generated as Poisson processes, with rate modulated by the pixel amplitude, and 100ms duration per data sample. 64 output neurons were simulated as standard leaky integrate-and-fire neurons, with $V_{rest} = -65mV$, $V_{threshold} = -50$, $V_{reset} = -65mV$, $\tau_{mem} = 15ms$. We simulated an input mean cancellation mechanism by subtracting the estimated input firing rate, with time scale $\tau_4 = 200s$, in the input current.

The minimal triplet-STDP model\cite{pfister_triplets_2006}
was implemented with weight decay and a homeostatic factor, in which synaptic changes follow 
\begin{equation}
\frac{d}{dt}w(t)=\eta^{+}y(t)\bar{y}^{+}(t)\bar{x}^{+}(t)-\eta^{-} h_y x(t)\bar{y}^{-}(t) - \lambda w(t)
\end{equation}
where $y(t)$ and $x(t)$ are the post- and pre-synaptic spike trains,
respectively: $y(t)=\sum_{f}\delta(t-t^{f})$, where $t^{f}$ are
the firing times and $\delta$ denotes the Dirac $\delta$-function;
$x(t)$ is a vector with components $x_{i}(t)=\sum_{f}\delta(t-t_{i}^{f})$,
where $t_{i}^{f}$ are the firing times of pre-synaptic neuron $i$. $\eta^{+}=10^{-4}$, $\eta^{-}=10^{-4}$ and $\lambda = 0.05$ are unit-free constants, and $\bar{y}^{+}$,
$\bar{x}^{+}$ and $\bar{y}^{-}$ are moving averages, implemented
by integration (e.g. $\tau\frac{\partial\bar{{y}}}{\partial t}=-\bar{y}+y$),
with time scales of $30$ ms. The homeostatic factor $h_y = \langle y \rangle^2$, estimated with a time scale $\tau_h = 200s$. The Oja-like STDP model was composed of the triplet LTP factor and a heterosynaptic LTD factor, 
\begin{equation}
\frac{d}{dt}w(t)=\eta^{+}y(t)\bar{y}^{+}(t)\bar{x}^{+}(t) - \eta^{-} w(t) h_y
\end{equation}
with $\eta^{+}=10^{-4}$, $\eta^{-}=10^{-4}$ and $h_y = \langle y \rangle^2$, estimated with a time scale $\tau_h = 200s$.

Recurrent inhibitory plasticity was adapted from \citep{vogels_inhibitory_2011}, with weight decay, with synaptic changes following
\begin{equation}
\frac{d}{dt}w(t)=\eta \ \left( \bar{x}(t)(y(t)-\theta) + x(t)(\bar{y}(t)-\theta) \right) - \lambda w(t)
\end{equation}
with constants $\eta = 0.001$, $\theta = 0.003$ and $\lambda = 3.0$. 


\nolinenumbers


 
\printbibliography
  
\end{document}